\documentclass[12pt]{iopart}
\usepackage{amsfonts}
\usepackage[english]{babel}
\usepackage{euscript}
\usepackage{bbm}

\begin{document}

\title[B\"acklund transformations between four Lax-integrable 3D equations]%
{B\"acklund transformations between four Lax-integrable 3D equations}

\author{Oleg I. Morozov$^1$}

\address{$^1$Faculty of Applied
  Mathematics, AGH University of Science and Technology,
  \\
  Al. Mickiewicza 30,
  Krak\'ow 30-059, Poland
\\
morozov{\symbol{64}}agh.edu.pl}

\author{Maxim  V. Pavlov$^{2,3,4}$}

\address{$^{2}$ Sector of Mathematical Physics, Lebedev Physical Institute of Russian Academy of Sciences,
         Leninskij Prospekt 53, 119991 Moscow, Russia;}
\address{$^{3}$ Department of Applied Mathematics, National Research Nuclear University MEPHI, Kashirskoe
         Shosse 31, 115409 Moscow, Russia;}
\address{$^{4}$ Department of Mechanics and Mathematics, Novosibirsk State University, 2 Pirogova street,
         630090, Novosibirsk, Russia}

\begin{abstract}
We show that four Lax-integrable 3D differential equations are related via B\"acklund transformations.
\end{abstract}



\maketitle

The aim of this note is to construct B\"acklund transformations,
\cite{KrasilshchikVinogradov1989,VinogradovKrasilshchik1997},
 between the following four equations
\begin{equation}
u_{yy} = u_{tx} + u_y \,u_{xx} - u_x\,u_{xy},
\label{Pe}
\end{equation}
\begin{equation}
u_{ty} = u_x\,u_{xy} - u_y \,u_{xx},
\label{Oe}
\end{equation}
\begin{equation}
u_{yy} = u_y\,u_{tx} - u_x \,u_{ty},
\label{uhe}
\end{equation}
\begin{equation}
u_{ty} = u_t\,u_{xy} - u_y \,u_{tx},
\label{mVwe}
\end{equation}
In different contexts, these equations have been appeared before in
\cite{
Blaszak2002,%
Pavlov2003,%
Dunajski2004,%
MartinezAlonsoShabat2004,%
Pavlov2006,%
AdlerShabat2007,%
ManakovSantini2009,%
OdesskiSokolov2010,%
Ovsienko2010,%
Morozov2009,%
Morozov2012a,%
Morozov2012b,%
FerapontovMoss2015,%
BKMV2016}. The equations are Lax-integrable, that is, each
equation has a differential covering (or a Lax pair) with a non-removable spectral parameter. The coverings for
(\ref{Pe}), (\ref{Oe}), (\ref{uhe}), (\ref{mVwe}), are defined,
\cite{Pavlov2003,Dunajski2004,Pavlov2006,Morozov2009,MartinezAlonsoShabat2004,AdlerShabat2007},
by the following over-determined systems,
respectively:
\begin{equation}
\left\{
\begin{array}{lcl}
v_t &=& (\lambda^2-\lambda\,u_x-u_y)\,v_x,
\\
v_y &=& (\lambda-u_x)\,v_x,
\end{array}
\right.
\label{Pe_covering}
\end{equation}
\begin{equation}
\left\{
\begin{array}{lcl}
v_t &=& (u_x-\lambda)\,v_x,
\\
v_y &=& \lambda^{-1}\,u_y\,v_x,
\end{array}
\right.
\label{Oe_covering}
\end{equation}
\begin{equation}
\left\{
\begin{array}{lcl}
v_t &=& \lambda^{-1}\,u_y^{-1}\,v_y,
\\
v_x &=& (\lambda+u_y\,u_x^{-1})\,v_y,
\end{array}
\right.
\label{uhe_covering}
\end{equation}
\begin{equation}
\left\{
\begin{array}{lcl}
v_t &=& \lambda\,(\lambda+1)^{-1}\,u_t\,v_x,
\\
v_y &=& \lambda\,u_y\,v_x.
\end{array}
\right.
\label{mVwe_covering}
\end{equation}
Parameter $\lambda \in \mathbb{R}$ is assumed to satisfy conditions $\lambda \ne 0$ in (\ref{Oe_covering}),
(\ref{uhe_covering}) and $\lambda \not \in \{-1, 0\}$ in (\ref{mVwe_covering}). The compatibility conditions for
(\ref{Pe_covering}), (\ref{Oe_covering}), (\ref{uhe_covering}), and (\ref{mVwe_covering}) coincide with
equations (\ref{Pe}), (\ref{Oe}), (\ref{uhe}), and (\ref{mVwe}), respectively.
Excluding $u$ from (\ref{Pe_covering}), (\ref{Oe_covering}), (\ref{uhe_covering}) with $\lambda=1$ and from
(\ref{mVwe_covering}) for arbitrary $\lambda$ yields equations
\begin{equation}
v_{yy} = v_{tx} + \frac{v_y-v_t}{v_x} \,v_{xx} +\frac{v_y-v_x}{v_x}\,v_{xy},
\label{mPe}
\end{equation}
\begin{equation}
v_{ty} = \frac{v_t+v_x}{v_x}\,v_{xy} - \frac{v_y}{v_x} \,v_{xx},
\label{mOe}
\end{equation}
\begin{equation}
v_{yy} = \frac{v_y}{v_t}\,v_{tx} +\frac{v_y- v_x}{v_t} \,v_{ty},
\label{muhe}
\end{equation}
\begin{equation}
v_{ty} = \frac{\lambda+1}{\lambda}\,\frac{v_t}{v_x}\,v_{xy} - \frac{1}{\lambda}\,\frac{v_y}{v_x} \,v_{tx}.
\label{Vwe}
\end{equation}
In other words, systems (\ref{Pe_covering}), (\ref{Oe_covering}), (\ref{uhe_covering}), (\ref{mVwe_covering})
define B\"acklund transformations between pairs of equations (\ref{Pe}) and (\ref{mPe}), (\ref{Oe}) and
(\ref{mOe}), (\ref{uhe}) and (\ref{muhe}), (\ref{mVwe}) and (\ref{Vwe}), respectively.
Equation (\ref{Vwe}) was considered in \cite{Zakharevich2000}.

\newpage
\noindent
{\sc Theorem}. {\it The following pairs of equations are equivalent via point transformations:
\begin{itemize}
\item[(i)]
(\ref{mPe}) and (\ref{Oe}),
\item[(ii)]
(\ref{mOe}) and (\ref{muhe}),
\item[(iii)]
(\ref{muhe}) and (\ref{mVwe}).
\end{itemize}
}
\vskip 7 pt
\noindent
Proof. (i)  Write equation (\ref{mPe}) as
\begin{equation}
\tilde{v}_{\tilde{y}\tilde{y}} =
\tilde{v}_{\tilde{t}\tilde{x}}
+ \frac{\tilde{v}_{\tilde{y}}-\tilde{v}_{\tilde{t}}}{\tilde{v}_{\tilde{x}}} \,\tilde{v}_{\tilde{x}\tilde{x}}
+\frac{\tilde{v}_{\tilde{y}}-\tilde{v}_{\tilde{x}}}{\tilde{v}_{\tilde{x}}}\,\tilde{v}_{\tilde{x}\tilde{y}}.
\label{mPe_tilded}
\end{equation}
Then the change of variables
\begin{equation}
\tilde{t} = t,
\quad
\tilde{x} = -u+x,
\quad
\tilde{y} = x,
\quad
\tilde{v} = y
\label{PT1}
\end{equation}
maps equation (\ref{mPe_tilded}) to equation (\ref{Oe}).
\vskip 5 pt
(ii)  Write equation (\ref{muhe}) as
\begin{equation}
\tilde{v}_{\tilde{y}\tilde{y}} =
\frac{\tilde{v}_{\tilde{y}}}{\tilde{v}_{\tilde{t}}}\,\tilde{v}_{\tilde{t}\tilde{x}}
+\frac{\tilde{v}_{\tilde{y}}- \tilde{v}_{\tilde{x}}}{\tilde{v}_{\tilde{t}}} \,\tilde{v}_{\tilde{t}\tilde{y}}.
\label{muhe_tilded}
\end{equation}
Then the change of variables
\begin{equation}
\tilde{t} = y,
\quad
\tilde{x} = t,
\quad
\tilde{y} = -x,
\quad
\tilde{v} = v
\label{PT2}
\end{equation}
maps equation (\ref{muhe_tilded}) to equation (\ref{mOe}).
\vskip 5 pt
(iii)
The change of variables
\begin{equation}
\tilde{t} = t,
\quad
\tilde{x} = x,
\quad
\tilde{y} = u,
\quad
\tilde{v} = y
\label{PT3}
\end{equation}
maps equation (\ref{muhe_tilded}) to equation (\ref{mVwe}).
\hfill $\Box$
\vskip 7 pt
\noindent
{\sc Corollary}. {\it Each pair of equations
(\ref{Pe}), (\ref{Oe}), (\ref{uhe}), (\ref{mVwe}),
(\ref{mPe}), (\ref{mOe}), (\ref{muhe}), (\ref{Vwe}) is related via an appropriate combination
of transformations
(\ref{Pe_covering}), (\ref{Oe_covering}), (\ref{uhe_covering}), (\ref{mVwe_covering}),
(\ref{PT1}), (\ref{PT2}), (\ref{PT3}).
}

\section*{Acknowledgments}
OIM gratefully acknowledges financial support from the Polish Ministry
of Science and Higher Education.
MVP's work was partially supported by the grant of Presidium of RAS
``Fundamental Problems of Nonlinear Dynamics'' and by the RFBR grant 14-01-00012.

\end{document}